\documentclass{PoS}

\newcommand{\be}{\begin{equation}}
\newcommand{\ee}{\end{equation}}
\newcommand{\ba}{\begin{eqnarray}}
\newcommand{\ea}{\end{eqnarray}}
\newcommand{\bi}{\begin{itemize}}
\newcommand{\ei}{\end{itemize}}
\newcommand{\tr}{{\rm Tr\,}}
\newcommand{\re}{\mathop{\rm Re}}

\newcommand{\half}{{\textstyle\frac{1}{2}}}

\newcommand{\<}{\langle}
\renewcommand{\>}{\rangle}
\newcommand{\eq}{Eq.~}

\newcommand{\fig}{Fig.~}

\newcommand{\la}{\label}

\newcommand{\txts}{\textstyle}

\newcommand{\im}{\mathop{\rm Im}}

\newcommand{\as}{a_{\sigma}}
\newcommand{\at}{a_{\tau}}

\newcommand{\Nt}{N_{\tau}}

\newcommand{\hZDs}{\widehat Z_{{\rm d}\sigma}}
\newcommand{\hZDt}{\widehat Z_{{\rm d}\tau}}

\newcommand{\hZThs}{\widehat Z_{{\theta}\sigma}}
\newcommand{\hZTht}{\widehat Z_{{\theta}\tau}}

\title{Energy-momentum tensor correlators and viscosity }

\ShortTitle{Energy-momentum tensor correlators and viscosity }

\author{\speaker{Harvey Meyer}%
\\
      Center for Theoretical Physics\\
      Massachusetts Institute of Technology \\
      Cambridge, MA 02139 U.S.A.\\
        E-mail: \email{meyerh@mit.edu}}

\abstract{
Collective flow has been observed in heavy ion collisions,
with a large anisotropic component,
and ideal hydrodynamic calculations had significant successful in
describing the distribution of produced particles at the RHIC experiments.
In order to account for this near ideal fluid behavior, the
shear and bulk viscosity of the quark gluon plasma (QGP)
must be computed from first principles in a regime 
where the QGP is not weakly coupled.
In this talk I describe recent progress in computing
energy-momentum tensor correlators on the lattice from
which the viscosities can be extracted via Kubo formulas.
I also show how to cumulate information from several channels,
including at non-vanishing spatial momentum,
in order to best constrain the viscosities.
These methods should soon yield predictions
at the higher temperatures that will 
be explored at the LHC experiments.
}

\FullConference{The XXVI International Symposium on Lattice Field Theory \\
                 July 14 - 19, 2008\\
                 Williamsburg, Virginia, USA}

\usepackage{epsfig}
\usepackage{graphicx}
\usepackage{dcolumn}
\usepackage{bm}

\begin{document}
\section{Introduction}

Heavy ion collisions at RHIC have produced a rich phenomenology
-- see the contribution of R.~Pisarski at this conference --
which has revealed unexpected properties of the quark gluon plasma.
Here I will be concerned with the thermodynamic and transport 
properties of the plasma in the range of temperatures $T_c\leq T\leq4T_c$.
Hydrodynamics calculations~\cite{Kolb:2000fha,Huovinen:2001cy,Teaney:2000cw} 
successfully described the distribution of produced particles in heavy ion collisions 
at RHIC~\cite{Arsene:2004fa,Back:2004je,Adams:2005dq,Adcox:2004mh}. 
This early agreement between ideal hydrodynamics 
and experiment has been refined in recent times.
On the theory side, the dissipative effects of shear viscosity $\eta$
have been included  in full 3d hydrodynamics 
calculations~\cite{Romatschke:2007mq,Dusling:2007gi,Song:2007fn}
and the sensitivity to initial conditions quantitatively estimated~\cite{Luzum:2008cw}
for the first time.
On the experimental side, the elliptic flow observable $v_2$,
which is sensitive to the value of $\eta$ in units of entropy density $s$, 
is now corrected for non-medium-generated two-particle correlations~\cite{:2008ed}.
The conclusion that $\eta/s$ must be much smaller than unity
has so far withstood these refinements of heavy-ion phenomenology~\cite{Luzum:2008cw}.

On the theory side, 
it is therefore important to compute the QCD shear viscosity from first principles
to complete the picture.
Furthermore, since the heavy ion collision program at LHC will probe the quark-gluon
plasma at temperatures about a factor two higher, it is crucial to predict
the shear viscosity at $\approx3T_c$, and from there to predict the size of elliptic flow,
before experimental data is available. 

A small shear viscosity is a signature of strong interactions, as I will explain in
the next section;
strong interactions in turn require non-perturbative computational techniques.
In this talk I present the lattice calculation of the
thermal correlators of the energy-momentum tensor
in the Euclidean SU(3) pure gauge theory, and
discuss methods to extract the shear and bulk viscosity from them.
Computationally, the calculation is challenging enough without the inclusion of dynamical quarks,
and physically, the thermodynamic properties of the QGP not too close to $T_c$ 
do not depend sensitively on the flavor content~\cite{Karsch:2006sf}. 
In perturbation theory~\cite{Arnold:2003zc}, the ratio of shear viscosity to
entropy density is O($\frac{1}{\alpha_s^2}$ and 
there is only about $30\%$ difference between the pure gauge theory 
and full QCD~\cite{Moore:2004kp} at a fixed value of $\alpha_s$ 
($\eta/s$ is smaller in the pure gauge theory).
It should be appreciated that this is not much for a quantity which is infinite 
in the Stefan-Boltzmann limit, and the difference is actually reduced when comparing 
the pure gauge theory and full QCD at a common value of $T/T_c$.

\section{The physical significance of viscosity}
\begin{figure}
\begin{center}
\vspace{-0.5cm}
\psfig{file=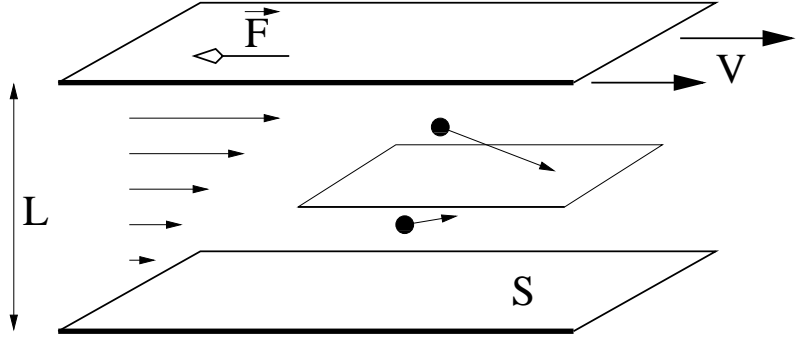,angle=0,width=10.5cm}
\caption{In the linear regime, the friction force per unit 
surface felt by a layer of fluid is proportional to the transverse velocity gradient.
The proportionality coefficient is the shear viscosity: $F/S=\eta V/L$.}
\la{fig:ideal}
\end{center}
\end{figure}

Consider a dilute gas enclosed in a box, \fig\ref{fig:ideal}. 
A vertical velocity gradient $V/L$
can be established by pulling the upper plate horizontally, 
at the cost of overcoming a friction force per unit surface $F/S$.
The macroscopic interpretation of that force is that neighboring layers of fluid 
are being dragged along. The microscopic mechanism is that 
molecules passing through a horizontal unit surface from above
carry more longitudinal momentum than those coming from below. 
This generates a net transfer of momentum in the transverse direction.
The longitudinal momentum of a molecule that passes through the unit
surface is characteristic of the average longitudinal momentum
in a slice located at a separation of order the mean free path 
$\ell_{\rm mfp}$ from the reference slice. 
The downward flux of molecules is clearly proportional
to their density  $n$. However the relation between the mean free path
and the density, $\ell_{\rm mfp}=\frac{1}{n\sigma}$, 
cancels that dependence in the net momentum flux
($\sigma$ is the interaction cross-section).
The viscosity is then given by $\eta=\frac{\<p\>}{3\sigma}$, where 
$\<p\>$ is the average momentum. The result is independent 
of the density $n$, at leading order in a density expansion.


A finite bulk viscosity has the effect of reducing the pressure
of an expanding fluid, as compared to the adiabatic expansion,
$P=P_{\rm equil.}-\zeta \nabla\cdot v$~\cite{Moore:2008ws}. For a given equation of state,
it can thus reduce the amount of radial flow in heavy ion collisions.
It can also play an important role in cosmology~\cite{Weinberg:1971mx}.
In a dilute gas with two-body interactions, the bulk viscosity is proportional 
to the density of molecules, unlike the shear viscosity, 
and is therefore much smaller than the latter.

For further qualitative discussions of viscosity, see~\cite{Kapusta:2008vb,Moore:2004kp}.
%

\section{Methodology}
The set of operators that plays a central role in the calculation of the bulk and shear viscosities
is the energy-momentum tensor.
The Euclidean energy-momentum tensor for SU($N_c$) gauge theories reads
\ba
T_{\mu\nu}(x) &=& \theta_{\mu\nu}(x) + \frac{1}{4}\delta_{\mu\nu}\,\theta(x)
\\
\theta_{\mu\nu}(x) &=&
 {\txts\frac{1}{4}}\delta_{\mu\nu}F_{\rho\sigma}^a F_{\rho\sigma}^a
   - F_{\mu\alpha}^a F_{\nu\alpha}^a
\\
\theta(x) &=& {\txts\frac{\beta(g)}{2g}} ~ F_{\rho\sigma}^a  F_{\rho\sigma}^a
\\
\beta(g) &=&  -b_0g^3+\dots,\qquad  b_0={\txts\frac{11N_c}{3(4\pi)^{2}}}\,.
\ea
In contrast with Minkovsky space,  $T_{0i}=\theta_{0i}$ is an antihermitian operator.
In particular, $\<T_{0i}(x)T_{0i}(0)\><0$ for $x\neq0$, and $P_j = i\int d^3{\bf x}\, T_{0j}(x) $
is the usual momentum operator, for instance $P_j|{\bf q}\> = q_j|{\bf q}\>$
for a one-particle state.

The Euclidean correlators to be computed on the lattice are ($L_0=1/T$)
\be
C_{\mu\nu,\rho\sigma}(x_0,{\bf q},T)= L_0^5\int d^3{\bf x} ~e^{i{\bf q\cdot x}} 
                             ~\< T_{\mu\nu}(0) T_{\rho\sigma}(x_0,{\bf x}) \>\,.
\ee
For diagonal correlators, the spectral functions are then defined by
\ba
C_{\mu\nu,\mu\nu}(x_0,{\bf q},T) = L_0^5 \int_0^\infty \rho_{\mu\nu,\mu\nu}(\omega,{\bf q},T) 
\,\frac{\cosh \omega(\half L_0-x_0)}{\sinh \frac{\omega L_0}{2}}\, d\omega\,.
\la{eq:int_rho}
\ea
(no summation over $\mu,\nu$ is implied here). The spectral functions are positive, 
$\rho(\omega,{\bf q})/\omega\geq0$, and odd, $\rho(-\omega,{\bf q})=-\rho(\omega,{\bf q})$. 
It is straightfoward to obtain a spectral 
representation for $\rho$, and thereby to show~\cite{Meyer:2008dq}
that it is related to the Minkovsky-space retarded correlator via
$\rho(\omega,{\bf q},T) = -\frac{1}{\pi}\im G_R(\omega,{\bf q},T)$.
The Kubo formulas then relate the low-frequency behavior of $G_R(\omega,{\bf q},T)$
with the shear and bulk viscosity, see~\cite{Teaney:2006nc} for a derivation.
The shear and bulk viscosities are given by~\cite{Karsch:1986cq}
\be
\eta(T) = \pi \lim_{\omega\to0} \frac{\rho_{12,12}(\omega,{\bf 0},T)}{\omega},\qquad\qquad
\zeta(T) = \frac{\pi}{9}\sum_{k,\ell} \lim_{\omega\to0} 
\frac{\rho_{kk,\ell\ell}(\omega,{\bf 0},T)}{\omega} .
\ee

In Ref.~\cite{Meyer:2007dy}, 
I defined the following moments of the spectral function ($n=0,1,\dots$):
\be
 \<\omega^{2n}\>\equiv L_0^5\int_0^\infty d\omega
   \frac{\omega^{2n}\rho(\omega)}{\sinh \omega L_0/2}
       = \left.\frac{d^{2n}C}{dx_0^{2n}}\right|_{x_0=L_0/2}
\la{eq:moments}
\ee
The latter equality implies that they are directly accessible to  lattice calculations.

\section{Discretization of $T_{\mu\nu}$}
In the following sections I present results obtained with the `clover'
discretization of the energy-momentum tensor~\cite{Gockeler:1996mu,Meyer:2007ic}.
I will use the  anisotropic Wilson gauge actions, 
see for instance~\cite{Namekawa:2001ih}.
Indeed it has long been recognized that such a lattice presents advantages
for the calculation of thermodynamics~\cite{Namekawa:2001ih} 
and thermal correlation functions,
in particular in charmonium calculations~\cite{Asakawa:2003re}.
The downside is that more normalization factors are required.

The field strength tensor $\widehat F_{\mu\nu}(x)$ is given by
\ba
\widehat F_{0k}&=& \frac{1}{8\as\at}\left(Q_{0k}(x)-Q_{k0})\right) \\
\widehat F_{kl} &=& \frac{1}{8\as^2} \left(Q_{kl}(x)-Q_{lk}(x)\right).
\ea
It is the obvious generalization to the anistropic lattice
of the definition found in~\cite{Luscher:1996sc}. We define 
the normalized operators as follows,
\ba
\as^3\at\theta_{00}(x) &=&  \frac{1}{g_0^2}\re\tr\Big\{
\xi_0 \hZDt(g_0,\xi_0)\sum_{k} (\as^2\at^2\widehat F_{0k}(x)^2)
-\xi_0^{-1}\hZDs(g_0,\xi_0)\sum_{k<l}  (\as^4\widehat F_{kl}(x)^2)\Big\},
\nonumber\\
\as^3\at\theta(x)  &=&  -\frac{dg_0^{-2}}{d\log \as}\re\tr\Big\{
\xi_0 \hZTht(g_0,\xi_0)\sum_{k} (\as^2\at^2\widehat F_{0k}(x)^2)
+\xi_0^{-1}\hZThs(g_0,\xi_0)\sum_{k<l}  (\as^4\widehat F_{kl}(x)^2)\Big\},
\nonumber\\
\as^3\at\theta_{03}(x) &=& \frac{2\widehat Z_{\bf p}}{g_0^2}
 \re\tr\Big[(\as^3\at\widehat F_{01}\widehat F_{31}) + 
(\as^3\at\widehat F_{02} \widehat F_{32}) \Big]
\ea
Here $\xi_0$ is the bare anisotropy, and $\xi$ the renormalized one.
In total,  $\theta$ requires two normalization factors,
and $\theta_{\mu\nu}$ requires five of them.

To fix the normalization factors of $\theta_{00}$ and $\theta$ 
we  proceed as follows. Let $L$ be a physical length.
Choose an ensemble `1' with $L_0=L_1=L_2=L_3=L$ and an ensemble `2' 
with $L_0/2=L_1=L_2=L_3=L$;
then $F_0 \equiv L^4(\<\theta_{00}\>_2 - \<\theta_{00}\>_1)$ is an RGI quantity.
The subtraction is necessary on the anisotropic lattice
in order to cancel the mixing with the unit operator.
Consider now an ensemble `3' where $L_0=L_2=L_3=L_1/2=L$,
and compute $F_1\equiv L^4(\<\theta_{11}\>_3 - \<\theta_{11}\>_1)$.
By symmetry between direction $\hat 1$ and $\hat 0$, $F_0=F_1$.
This condition fixes the ratio of the two renormalization factors.

Similarly, requiring the equality of the RGI quantities 
$G_0\equiv L^4( \<\theta\>_2 - \<\theta\>_1 )$
and $G_1\equiv L^4( \<\theta\>_3 - \<\theta\>_1 )$ fixes the relative size of
the two normalization factors entering the trace anomaly operator $\theta$.

To fix the absolute size of the normalization factors,
we equate the clover-discretized conformality measure $\<\theta\>$ 
with the plaquette discretization, for which the 
normalization factors are known in terms of the lattice beta function~\cite{Namekawa:2001ih}
and the renormalization of the anisotropy~\cite{Klassen:1998ua}.
For the case of $\theta_{00}$, we obtain its absolute normalization
by calibrating its expectation value to the value of the entropy density obtained 
in~\cite{Namekawa:2001ih} from the integral method.

The normalization factor of $T_{0k}$ is determined by requiring 
that $C_{03,03}(x_0=1/2T,{\bf 0},T)=s/T^3$. This equality can be checked 
explicitly in hydrodynamics and in free field theory.

\subsection{Cutoff effects in perturbation theory}
The cutoff effects affecting the Euclidean correlators can be 
studied, and, eventually, removed order by order in perturbation theory.

In the continuum and infinite volume limit, the two-point function of 
$\half(T_{11}-T_{22})$ equals the two-point function of $T_{12}$
for all ${\bf q}=(0,0,q)$.  On the infinite lattice, 
discretizing either form yields
a correlator that approaches the continuum limit with O($a^2$) discretization errors.
Figure~\ref{fig:a2-effects} shows a comparison of the discretization errors
affecting these two discretization schemes at $\xi=1$.
Using the clover discretization of $T_{\mu\nu}$
at a given value of $\Nt$, discretizing  $T_{12}$
yields smaller discretization errors than discretizing $\half(T_{11}-T_{22})$,
for all values of $x_0$.
The figure also shows
that the cutoff effects of the plaquette discretization used in~\cite{Meyer:2007ic}
are almost identical to those obtained with the clover discretization of $T_{12}$.

In order to find the optimal range of anisotropies, we consider
the cutoff effects on the tensor correlators at a fixed value of the
spatial lattice spacing, $\as/L_0=$fixed. We then vary the anisotropy
between 1 and 4. On the right panel of \fig\ref{fig:a2-effects}, we see that the
sign of the cutoff effects changes for $1<\xi<2$, and goes to a finite value
in the Hamiltonian limit, $\xi\to\infty$. It is clearly seen that any choice $\xi\geq2$
reduces the cutoff effects significantly. It also appears that choosing $\xi>3$ does
not reduce the cutoff effects further, because they are dominated by the coarseness
of the spatial discretization. The cutoff effects are minimal near $\xi=2$,
although, because of the sign change, this is partly accidental.

It should be noted that \eq\ref{eq:int_rho} holds in the continuum, and therefore
the Euclidean correlator should in principle first be extrapolated 
in the standard way, before one attempts to reconstruct the spectral function.
In practice, we often use data at finite lattice spacing instead,
but one must then be careful not to use data at too small $x_0/a$, 
where cutoff effects become large. In the data presented in section 6, we have
reduced cutoff effects by dividing the non-perturbative lattice correlator
by the treelevel lattice correlator and multiplying it by the continuum 
correlator.

\begin{figure}
\begin{center}
\begin{minipage}{7.5cm}
\psfig{file=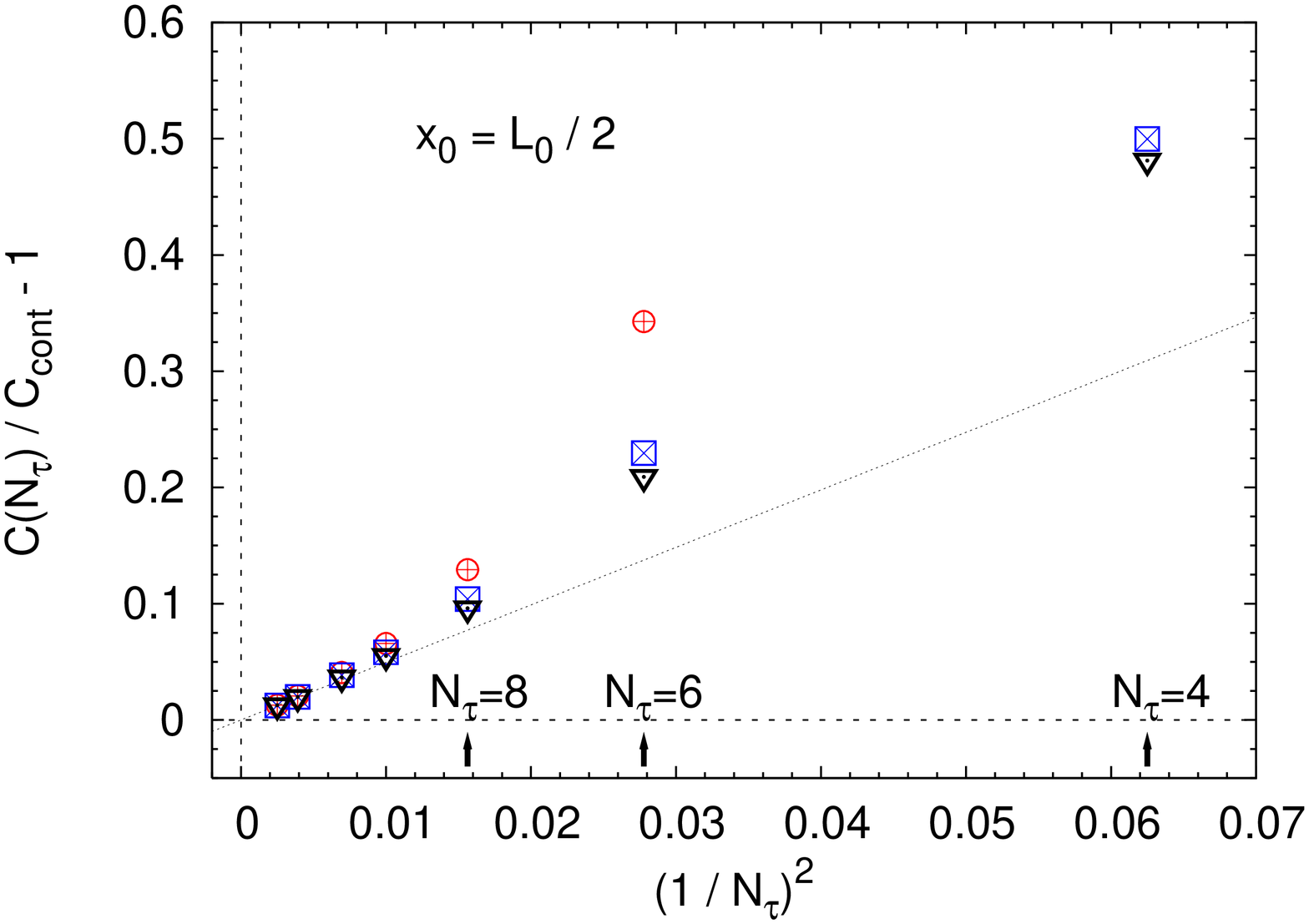 ,angle=-90,width=7.8cm}
\end{minipage}
\begin{minipage}{7.5cm}
\psfig{file=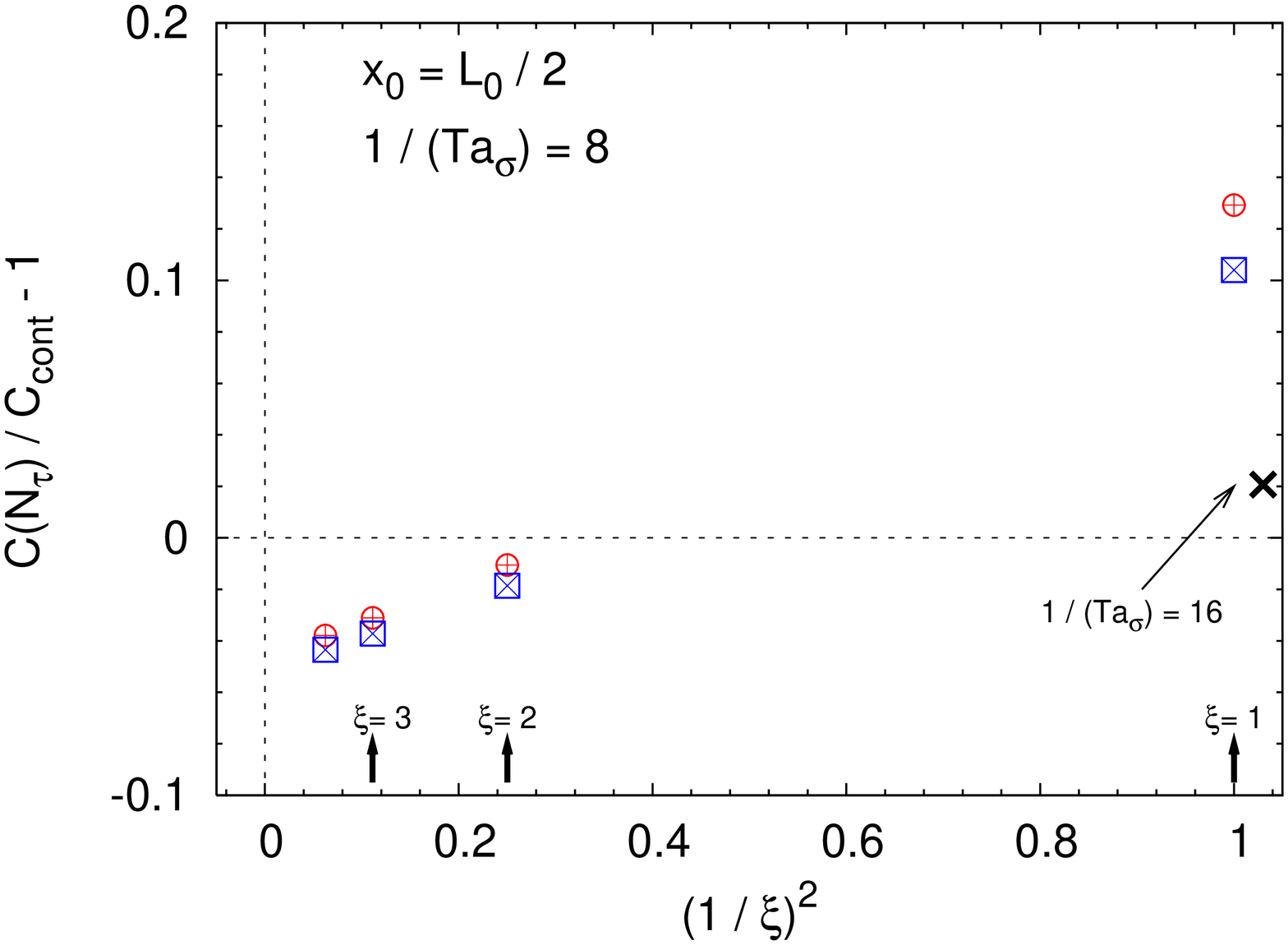,angle=-90,width=7.8cm}
\end{minipage}
\caption{\underline{Left}:  The relative deviation of the ${\bf p}=0 $
tensor correlator at finite lattice spacing from the continuum correlator.
The spatial volume is infinite.
The $\square$'s are for the clover discretization of $T_{12}$
and the $\otimes$ for the clover discretization 
of $\half(T_{11}-T_{22})$. The 
 $\nabla$ is the  plaquette discretization used in~\cite{Meyer:2007ic}.
\underline{Right}:  The cutoff effects on the tensor correlator for $x_0=L_0/2$. 
The spatial lattice spacing $\as$ is held fixed,
the temporal lattice spacing $\at$ is varied between a quarter and one times $\as$.
The $\square$'s refer to $T_{12}$ and the $\otimes$'s to $\half(T_{11}-T_{22})$.
For comparison, the cutoff effect on an isotropic $\Nt=16$  lattice
is indicated by a cross.}
\la{fig:a2-effects}
\end{center}
\end{figure}

\section{Analytic knowledge of the spectral function}
In this section we gather some of the known useful facts about the spectral functions.

\subsection{Ward identities}
The Ward identities that result from the conservation of 
the energy-momentum tensor, $\partial_\mu T_{\mu\nu}=0$, 
imply, for ${\bf q}=q\hat e_3$,
\ba
\omega^4\,\rho_{{00},{00}}(\omega,{\bf q})  = 
-\omega^2\,{q}^2\,\rho_{{03},{03}}(\omega,{\bf q})
&=& q^4 \rho_{33,33}(\omega,{\bf q})
\\
-\omega^2\,\rho_{{01},{01}}(\omega,{\bf q}) & =& q^2\,\rho_{13,13}(\omega,{\bf q})\,.
\la{eq:0202e1}
\ea
The first set of spectral functions is referred to as the `sound channel'
and the second as the `shear channel'~\cite{Teaney:2006nc}.
We therefore  use the notation 
$\rho_{\rm snd}(\omega,{\bf q})\equiv\rho_{{00},{00}}(\omega,{\bf q})$
and $\rho_{\rm sh}(\omega,{\bf q})\equiv  \rho_{{01},{01}}(\omega,{\bf q})$.

At zero temperature, Lorentz invariance imposes strong constraints
on the analytic form of the spectral function. This is 
shown in some detail in the Appendix for $\rho_{\rm snd}(\omega,{\bf q})$.
In particular, the Euclidean correlator $C_{00,00}$
vanishes as $|{\bf q}|^4$ at $T=0$, whereas it generically only 
vanishes as ${\bf q}^2$ at finite temperature.

\subsection{Weak coupling  predictions}
In~\cite{Meyer:2008gt}, I computed the spectral functions in 
the free field theory approximation. They correspond to 
 the result for photons, times the color factor $d_A=N_c^2-1$.
At finite momentum, they are functions of two dimensionless parameters,
$Tx_0$ and $q/T$, and are linear combinations of polylogarithms.
An example in the sound channel is shown on Fig.~\ref{fig:rho-analytic}.
Unlike in other channels illustrated in~\cite{Meyer:2008gt},
the spectral function has a discontinuity at threshold, $\omega=q$.
At zero-momentum, the spectral functions simplify considerably. 
For instance, 
\ba
\rho_{33,33}(\omega,{\bf 0},T) &=& 
\frac{2d_A}{15(4\pi)^2}\,\frac{\omega^4}{\tanh\frac{\omega}{4T}} 
+\left(\frac{2\pi}{15}\right)^2 3d_A T^4\,\omega \delta(\omega)\,,
\\
\rho_{12,12}(\omega,{\bf 0},T)&=& \frac{d_A}{10(4\pi)^2}\frac{\omega^4}{\tanh\frac{\omega}{4T}}
+\left(\frac{2\pi}{15}\right)^2 d_A T^4\,\omega \delta(\omega)\,.
\ea
These free-field theory  spectral functions of the fluxes
contain an $\omega\delta(\omega)$ term, which expresses the 
conservation of a momentum flux, an unrealistic feature 
from the QCD point of view. To describe the dissipation 
of momentum flux,  a more careful analysis is required, 
that was been carried out in~\cite{Aarts:2002cc,Moore:2008ws}.
These studies show that the delta function becomes a peak of finite width,
with a long tail in the $T_{kk}$ channel.

\begin{figure}
\begin{center}
\vspace{-0.5cm}
\psfig{file=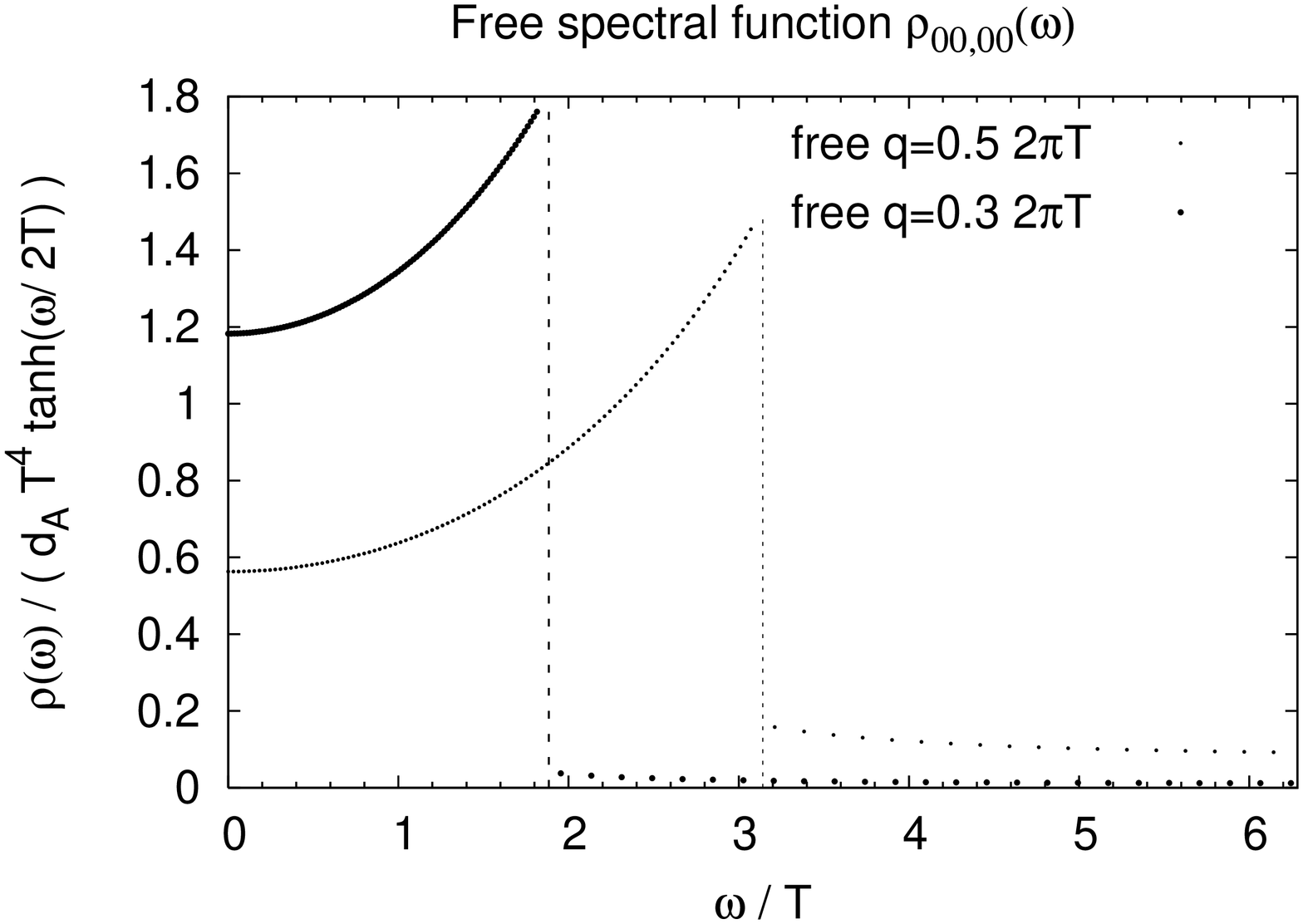,angle=-90,width=10.5cm}
\psfig{file=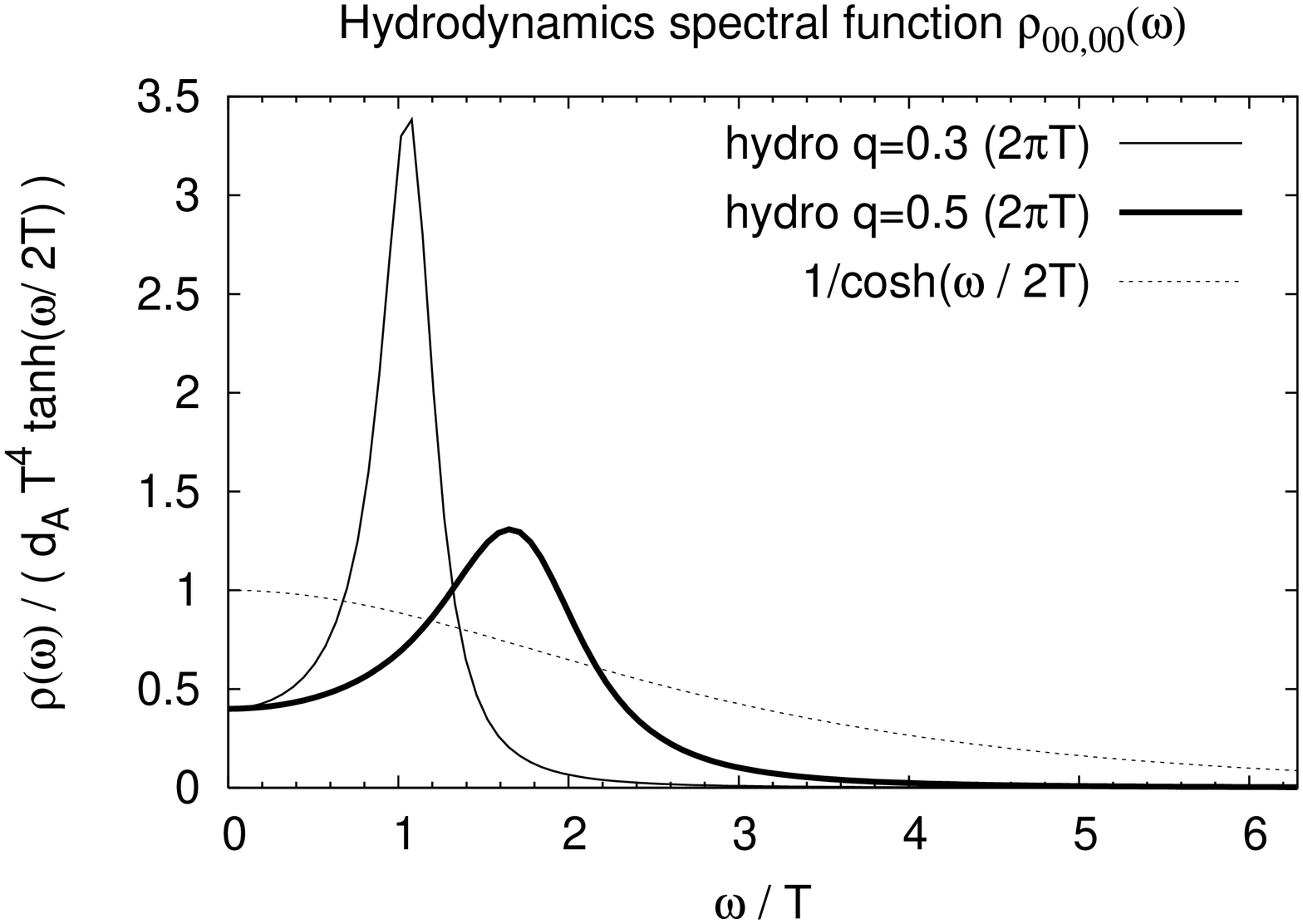 ,angle=-90,width=10.5cm}
.~~~~~~~~~~~~~~\psfig{file=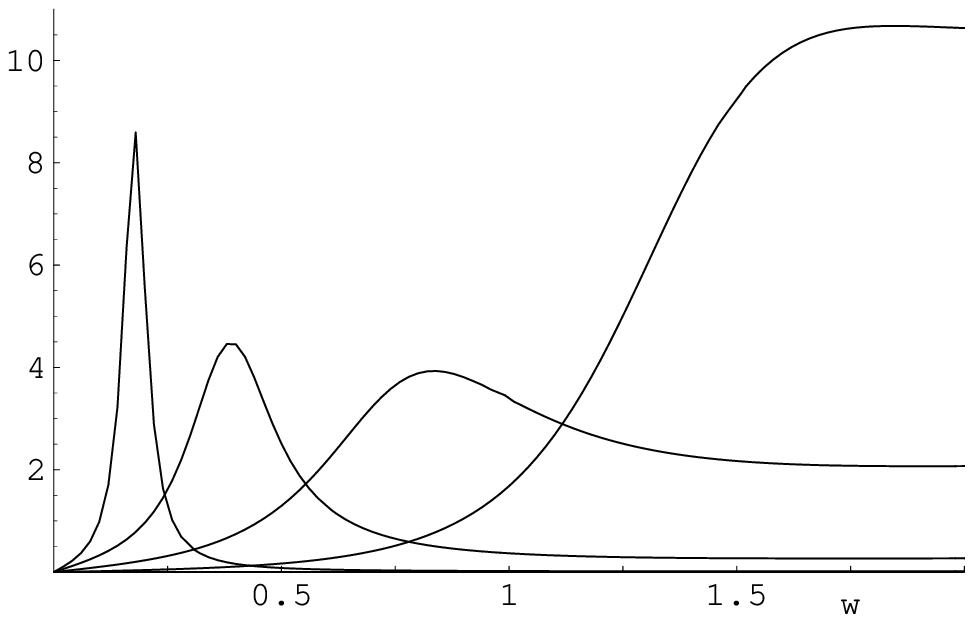,angle=0,width=9.8cm}
\caption{The $SU(N_c)$ gauge theory 
sound-channel spectral functions predicted by 
free theory (top). The functional form predicted by
 hydrodynamics (middle), for $v_s^2=\frac{1}{3}$, $s=\frac{3}{4}s_{SB}$
and $T\Gamma_s=\frac{1}{3\pi}$. 
At the bottom, the exact result for $2\rho/(\pi d_A T^4)$ 
in infinite-coupling, large-$N_c$ SYM 
theory, computed by AdS/CFT methods, 
is reproduced from~\cite{Kovtun:2006pf};
(the $x$-axis variable  is $w=\omega/2\pi T$ and 
the curves correspond to $q/2\pi T=0.3,0.6,1.0$ and 1.5).}
\la{fig:rho-analytic}
\end{center}
\end{figure}

\subsection{Predictions of hydrodynamics}
If the momentum ${\bf q}=(0,0,q)$ , the hydrodynamics predictions
for the shear and sound channel are respectively~\cite{Teaney:2006nc}
\ba
\frac{\rho_{\rm sh}(\omega,{\bf q})}{\omega}
&\stackrel{\omega,q\to 0}{\sim} & ~
\frac{\eta}{\pi} \frac{q^2}{\omega^2+(\eta q^2/(\epsilon+P))^2}\,,
\\
\frac{\rho_{\rm snd}(\omega,{\bf q})}{\omega}
 &\stackrel{\omega,q\to 0}{\sim}& ~  \frac{\Gamma_s}{\pi}
\frac{(\epsilon +P)\,q^4}{(\omega^2-v_s^2q^2)^2+(\Gamma_s\omega q^2)^2}\,,
\ea
where
\be
\Gamma_s = \frac{\frac{4}{3}\eta+\zeta}{\epsilon+P} 
\ee
is the sound attenuation length. The sound channel spectral function
is illustrated on \fig\ref{fig:rho-analytic}.

\subsection{Strongly coupled ${\cal N}=4$ super-Yang-Mills}
The spectral functions of the ${\cal N}=4$ $SU(N_c)$ super-Yang-Mills theory
were calculated by AdS/CFT methods in~\cite{Kovtun:2006pf,Teaney:2006nc}
at infinite 't Hooft coupling and large number of colors.
Those spectral functions are very smooth, and exhibit precisely 
the  behavior generically  predicted by hydrodynamics at low frequencies 
and momenta, as illustrated by \fig\ref{fig:rho-analytic}. 
The sound peak in  $\rho_{\rm snd}$ is the dominant
feature of the spectral function at least up to $q/2\pi T=0.6$.
This is encouraging for lattice calculations, since the data
presented in the next section already reach
non-zero momenta as low as $q/2\pi T=0.125$.

\section{Numerical results}
We simulate the $\xi=2$ anisotropic Wilson gauge action with 
the two-level algorithm as described in~\cite{Meyer:2002cd,Meyer:2003hy}.
We observe that  correlators  with a small, non-vanishing momentum 
are more accurate than at ${\bf q}=0$. The typical accuracy achieved 
on the correlators is $1\%$, and in some cases it is as good as 
$0.5\%$. Here we show results for the temperatures 1.27, 1.52, 2.52
and 3.15 in units of $T_c$, respectively from lattices $12\cdot48^3$,
$10\cdot20^3$, $10\cdot20^3$ and $8\cdot24^3$. 

\subsection{Sound channel}
\begin{figure}
\begin{center}
\vspace{-0.5cm}
\psfig{file=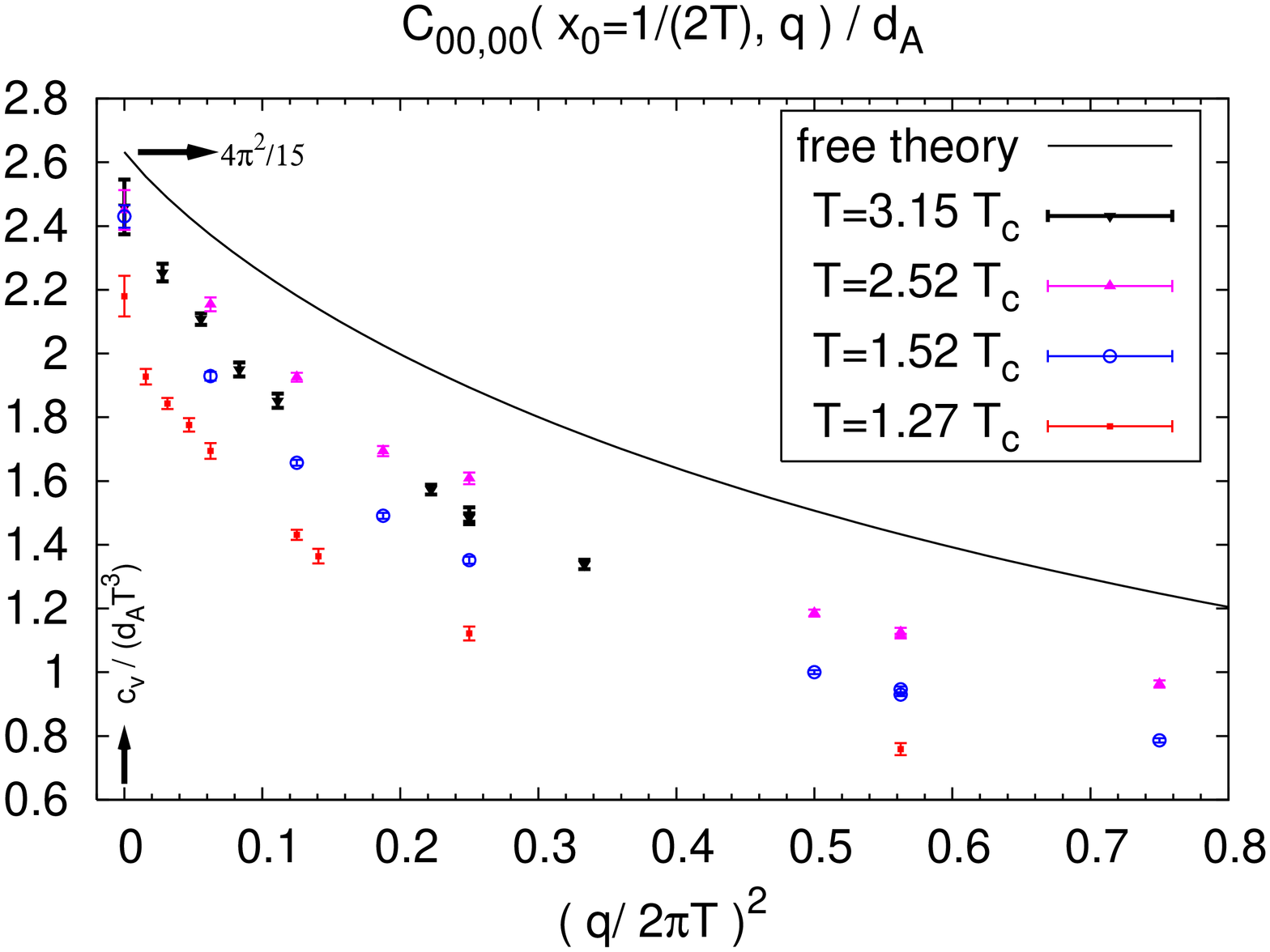,angle=-90,width=10.5cm}
\psfig{file=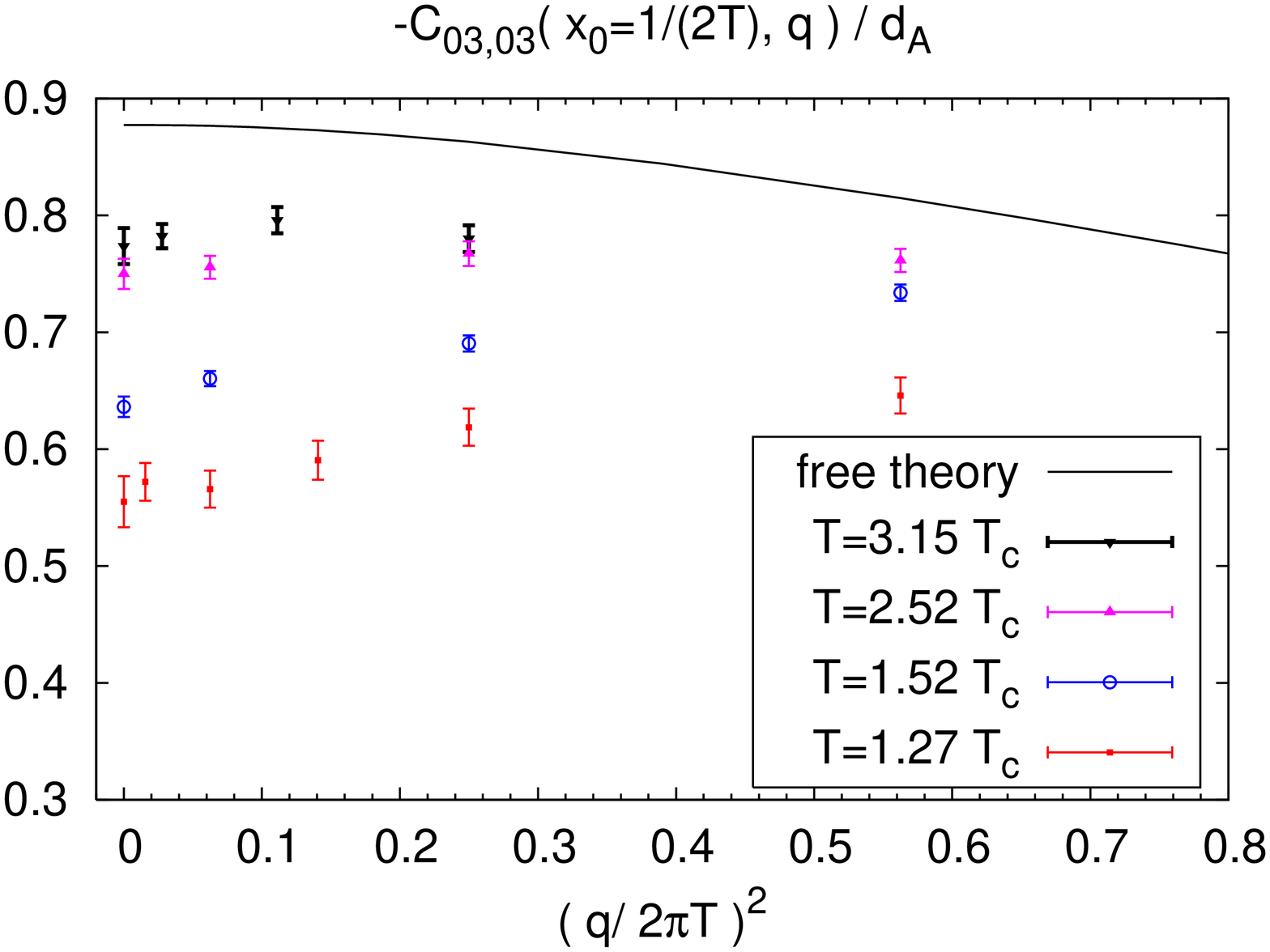,angle=-90,width=10.5cm}
\psfig{file=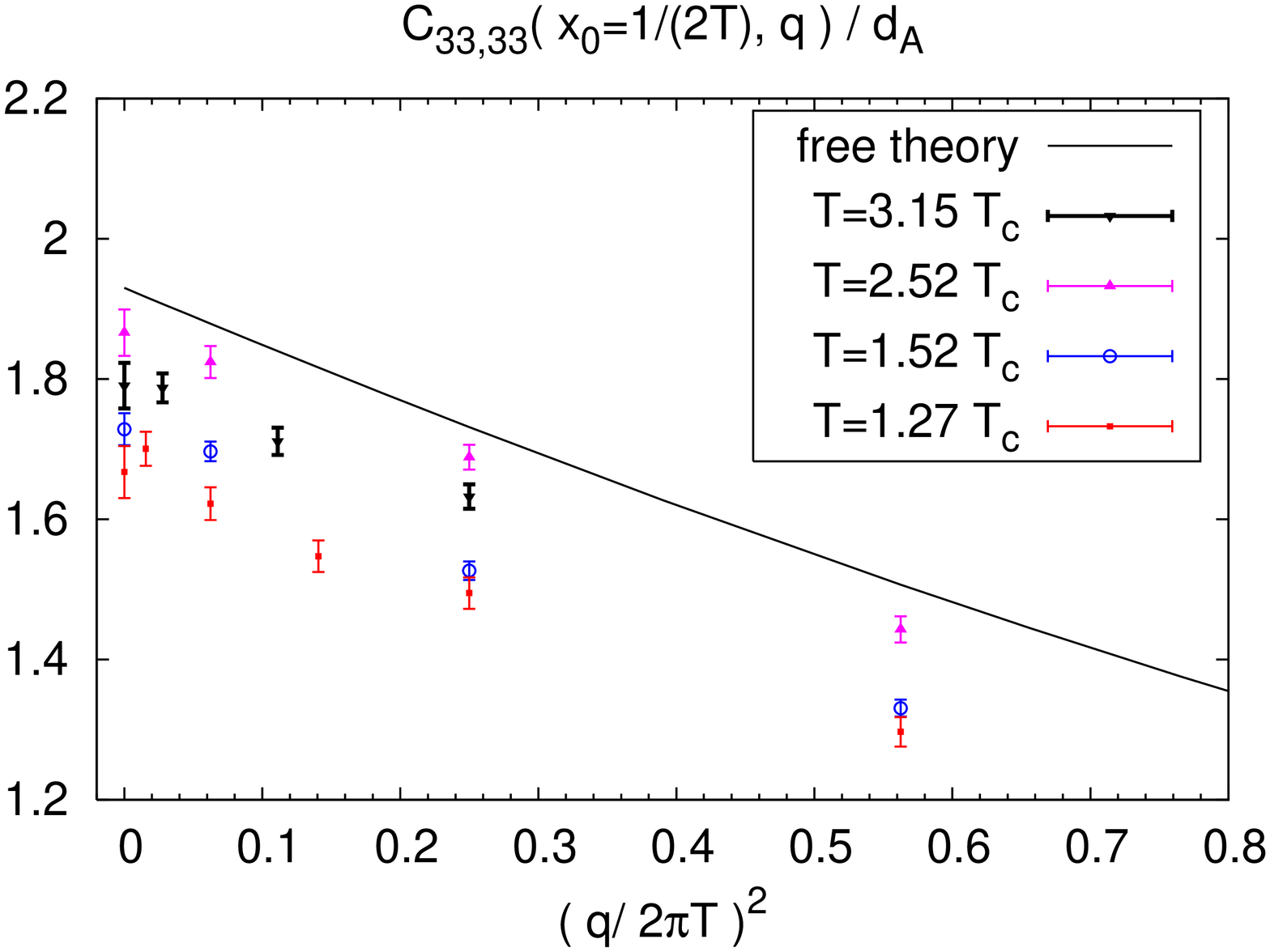,angle=-90,width=10.5cm}
\caption{The sound-channel Euclidean correlators at $x_0=1/2T$.}
\la{fig:Cqsound}
\end{center}
\end{figure}

Due to the Ward identities, the correlators are all determined 
by the spectral function of the energy density. In particular, 
at fixed $x_0=1/2T$, 
\ba
C_{00,00}(q,T) &=& \int_0^\infty d\omega\,
\frac{\rho_{\rm snd}(\omega,q,T)}{\sinh\omega/2T}\,,
\\
-C_{03,03}(q,T) &=& \frac{1}{q^2}\int_0^\infty d\omega\omega^2\,
\frac{\rho_{\rm snd}(\omega,q,T)}{\sinh\omega/2T}\,,
 \\
 C_{33,33}(q,T) &=& \frac{1}{q^4}\int_0^\infty d\omega \omega^4\,
 \frac{\rho_{\rm snd}(\omega,q,T)}{\sinh\omega/2T}\,.
\ea
Computing $C_{03,03}$ and $C_{33,33}$ thus provides complementary
information to $C_{00,00}$, without having to take finite differences 
in $x_0$, which necesarily leads to larger statistical and discretization errors.

Figure \ref{fig:Cqsound} displays the three correlators of the sound channel, 
as a function of spatial momentum ${\bf q}=q\hat e_3$.
One remarkable feature is that below $2T_c$, $|C_{03,03}|$ increases with $q^2$
at small momentum. This is unlike the free theory prediction.

An Ansatz that interpolates between the hydrodynamic behavior 
at low frequencies and the perturbative behavior at high frequencies is 
\be
\frac{\widehat\rho_{\rm snd}(\omega,q,T)}{\tanh(\frac{\omega}{2T})}=
\frac{\frac{2}{\pi}\,\Gamma_s\, (\epsilon+P)q^4}
{(\omega^2-v_s^2q^2)^2+(\Gamma_s\omega q^2)^2}\, (1-\tanh^2{\omega/2T})
+\frac{2d_A q^4}{15(4\pi)^2}\, \tanh^2{\omega/2T}.
\la{eq:ansatz}
\ee
Fitting to the correlator $C_{00,00}(\frac{1}{2T},{\bf q},T)$ at 
$1.27T_c$ for $q/(\frac{\pi T}{4})=1,\,\sqrt{2},\,\sqrt{3}$ and 2,  
I obtain $T\Gamma_s = 0.15(10)$. This is compatible with 
my previous estimates~\cite{Meyer:2007ic,Meyer:2007dy} for $\eta$ and $\zeta$.

\begin{figure}
\begin{center}
\psfig{file=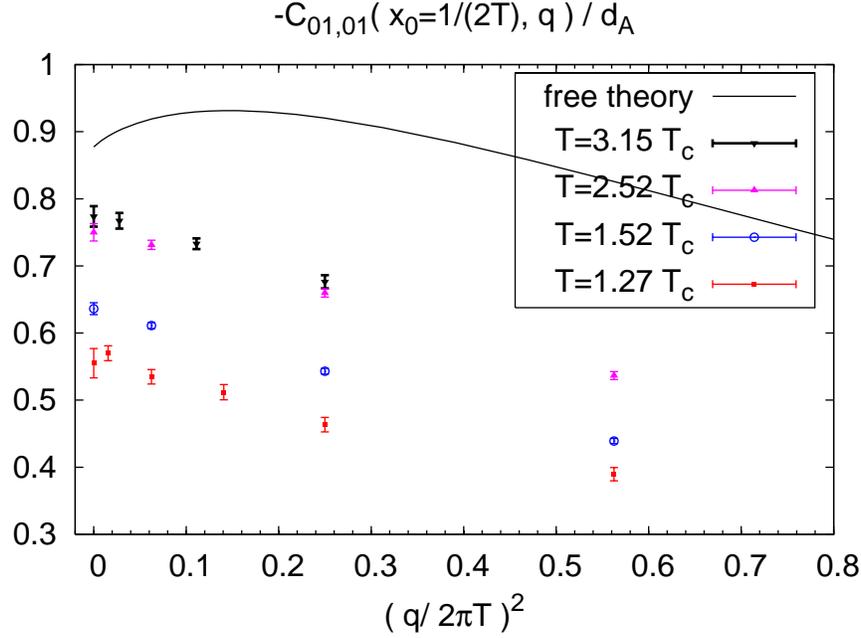,angle=-90,width=12.5cm}
\caption{A shear-channel Euclidean correlator at $x_0=1/2T$.}
\la{fig:Cqshear}
\end{center}
\end{figure}

In this (over-)simplified determination, 
I am relying  on the fact that the integral 
of the sound peak multiplied by $1/\cosh\frac{\omega}{2T}$ 
is of the form $c_vT^2+{\rm O}(q^2)$, 
and that the $q^2$ term has some dependence on
the width of the peak, itself given by the sound attenuation length. 
However, contributions from $\omega={\rm O}(T)$ are not parametrically suppressed,
although at these small momenta they are numerically very small 
both at weak coupling and in the strongly coupled SYM theory, see \fig\ref{fig:rho-analytic}.

A related point is that at order $q^2$, 
next-to-leading order hydrodynamic corrections to the sound
peak enter as well, and those are not described in Ansatz (\ref{eq:ansatz}).
For instance, the sound velocity has a momentum dependence at next-to-leading order.
At weak coupling~\cite{Moore:2008ws} 
 $\rho_{33,33}(\omega,{\bf 0},T)/\omega$ has a peak whose width $\omega_0$
is related to the inverse of a relaxation time~\cite{Karsch:2007jc}.
Since $\rho_{\rm snd}= (q/\omega)^4\rho_{33,33}$, this  effectively makes
the sound peak narrower, typically like
 $\frac{1}{(\Gamma_s q^2)^2}\to \frac{1}{(\Gamma_s q^2)^2} + \frac{1}{\omega_0^2}$.

Further progress will come from simultaneously analyzing the $x_0$ and 
$q$-dependence of the  correlator.

\section{Shear channel}

Similarly to the sound channel, at fixed $x_0=1/2T$
and with spatial momentum ${\bf q}=q\hat e_3$, 
\ba
-C_{02,02}(q,T) &=& \int_0^\infty d\omega\,
\frac{\rho_{\rm sh}(\omega,q,T)}{\sinh\omega/2T}\,,
\\
C_{12,12}(q,T) &=& \frac{1}{q^2}\int_0^\infty d\omega\,\omega^2\,
\frac{\rho_{\rm sh}(\omega,q,T)}{\sinh\omega/2T}\,.
\ea
Figure \ref{fig:Cqshear} displays the momentum correlator of the shear channel, 
as a function of spatial momentum. The derivative 
$\frac{d}{dq^2}C(\frac{1}{2T},q=0,T)$ is negative, unlike the 
free theory prediction. 

\section{Discussion}
\begin{figure}[t]
\begin{center}
\vspace{-1cm}
\psfig{file=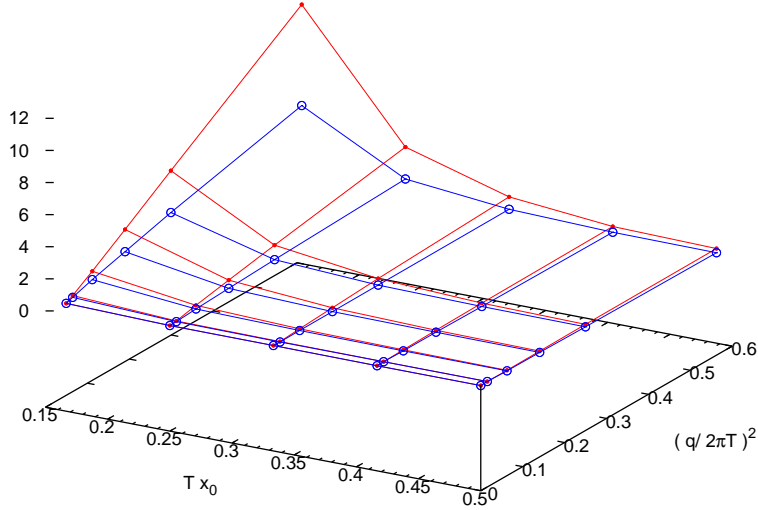,angle=-90,width=12.5cm}
\caption{Momentum and Euclidean time dependence  of the momentum density
correlator $C_{01,01}/d_A$ and $C_{03,03}/d_A$ 
in the shear and sound channels, corresponding to open and closed 
circle respectively. Error bars are smaller than the symbols.
The temperature is $1.27T_c$.}
\la{fig:3d}
\end{center}
\end{figure}
In the two-point functions of the momentum densities $T_{0k}$, 
I found that below $2T_c$ the first moment, 
$\frac{d}{dq^2}C_{0k,0k}(\frac{1}{2T},q=0,T)$,
is positive in the sound-channel ($k=3$), 
and negative in the shear channel $k=1$.
In free field theory, these signs are reversed. 
There is thus a qualitative difference between
these non-perturbative correlators and the correlators in the 
weak coupling limit. By contrast, the correlation functions 
of the fluxes at vanishing spatial momentum are numerically
not very different from the free-field 
prediction~\cite{Meyer:2007ic,Meyer:2007dy,Meyer:2008dq}.
This is because they are dominated by the perturbative $\omega^4$
term in the spectral function at high frequency.

I pointed out the advantage of the conserved-charge 
correlators in terms of sensitivity to the low-frequency region
$\omega\leq T$ region~\cite{Meyer:2008gt}.
More generally,  it is advantageous to constrain simultaneously the 
$\omega$ and ${\bf q}$ dependence of the spectral functions,
since hydrodynamics makes a combined prediction in these
two variables. Figure~\ref{fig:3d} displays the simultaneous
$x_0$ and $q$ dependence of the momentum density correlators
in the shear and sound channels. 
Thanks in part to treelevel improvement, 
the ${\bf q}=0$ correlator is $x_0$-independent to very good
accuracy, as it should be.
It is by fitting the whole surface parametrized by $(x_0,q)$ 
that one can best exploit the lattice data.

Recently, I have studied spatial correlation functions
of the energy-momentum tensor~\cite{Meyer:2008dt}. 
The advantage of these 
observables is that no analytic continuation is required:
they directly contain information on the space-like 
correlations present in the quark-gluon plasma.
They can also be related to the spectral functions:
\be
C({\bf r},T)= \lim_{\epsilon\to0}\int \frac{d{\bf q}}{(2\pi)^3}e^{i{\bf q\cdot r}}
\int_0^\infty d\omega e^{-\epsilon \omega}
\frac{\rho(\omega,{\bf q},T)}{\tanh\omega/2T},
\ee
which illustrates the wealth of information encoded in $\rho(\omega,{\bf q},T)$.
A hallmark of ordinary liquids is the short-range order 
observed (typically by neutron scattering) in the particle pair correlation
function, which displays several gradually damped oscillations.
One of the conclusions I reached is that the fluids with the smallest
shear viscosity are those that are dense, but that are not
ordered too strongly. The thermal contribution
to the energy density correlations exhibits a non-monotonic
behavior on distance scales of order $1/T$ that is markedly stronger than 
in leading order perturbation theory.
This may be the signature of a spatial ordering that is the quantum 
relativistic analogue of the short-range order characteristic of classical 
non-relativistic fluids.
An attractive strategy to test the claimed liquid-like properties of 
the quark-gluon plasma~\cite{Hirano:2005wx} 
is to compare these spatial correlators
with those computed in the strongly coupled SYM theory, which is 
known (from its spectral functions and small viscosity) to be 
an excellent fluid.

In conclusion, there has been much technical progress
since the first calculation of the Euclidean correlators relevant 
to shear and bulk viscosity~\cite{Karsch:1986cq,Nakamura:2004sy,Meyer:2007ic,Meyer:2007dy}. 
The first accurate data in the shear channel allowed me to derive
an upper bound on the shear viscosity, $\eta/s<1.0$ at $T<2T_c$
under only weak assumptions~\cite{Meyer:2007ic}.  To improve on this,
the current emphasis is on exploiting simultaneously all
the relevant lattice data, as well as the analytic properties 
of the spectral function known from perturbation theory in the 
ultraviolet and from hydrodynamics in the infrared.
Spatial correlators provide complementary information 
that may help us understand the ability of the strongly 
coupled QGP to flow.

\acknowledgments{
I thank Krishna Rajagopal and John Negele for valuable comments  and encouragement.
I thank Gert Aarts, Philippe de Forcrand, Berndt M\"uller and Derek Teaney for discussions.
The $\Nt=12$ simulations were performed on the BlueGene/L at MIT;
I thank Andrew Pochinsky for his assistance in running it efficiently.
The other simulations were performed on the desktop machines of 
the Laboratory for Nuclear Science, MIT.
This work was supported in part by funds provided by the 
U.S. Department of Energy under cooperative research agreement
DE-FG02-94ER40818.
}

\appendix

\section{The vacuum energy density correlator}
At zero temperature, we can exploit Lorentz invariance to constrain the analytic form 
of the spectral function. Following~\cite{Pivovarov:1999mr},
 \ba
\int d^4x e^{iqx}~ \<T_{\mu\nu}(x) T_{\rho\sigma}(0)\>
 &=& \eta_{\mu\nu,\rho\sigma}(q) T(q\cdot q)+f_{\mu\nu,\alpha\beta}(q) T_s(q\cdot q)
\\
 \eta_{\mu\nu}&=&  q_\mu q_\nu -q^2 \delta_{\mu\nu}, 
\qquad
f_{\mu\nu,\rho\sigma} = \eta_{\mu\nu}\eta_{\rho\sigma}
\\
\eta_{\mu\nu,\rho\sigma}&=&
 \eta_{\mu\rho}\eta_{\nu,\rho} +\eta_{\mu\sigma}\eta_{\nu\rho}
                            -\frac{2}{3} \eta_{\mu\nu}\eta_{\rho\sigma}.
\ea

This implies the following form  of the
energy density correlator in the mixed representation:
\be
C_{00,00}(x_0,{\bf q}) = |{\bf q}|^4\int_{-\infty}^\infty \frac{dq_0}{2\pi}
 e^{-iq_0x_0}[{\txts\frac{4}{3}} T(q_0^2+{\bf q}^2)+T_s(q_0^2+{\bf q}^2)].
\ee
In particular, the correlator satisfies
\be
\frac{\partial}{\partial x_0}
\left(\frac{\partial}{\partial |{\bf q}|}-\frac{4}{|{\bf q}|}\right)
C_{00,00}(x_0,{\bf q}) = |{\bf q}|x_0 \,C_{00,00}(x_0,{\bf q}).
\ee

On the other hand, if
\be
  v_n({\bf q}) \equiv \<n,{\bf q}|L^{-3/2}\int d^3{\bf x}\, e^{i{\bf q\cdot x}}
                 \widehat T_{00}({\bf x})|\Omega\>,
\ee
the K\"allen-Lehmann representation of the correlator reads
\be
C_{00,00}(x_0,{\bf q}) = \sum_n |v_n|^2({\bf q}) \, e^{-E_n({\bf q})x_0}. 
\ee
On then easily shows that  $\forall n$, 
\be
E_n^2({\bf q})  = E_n^2(0)+{\bf q}^2,  \qquad
|v_n|^2({\bf q}) = \frac{E_n^2(0) a_n^2 |{\bf q}|^4}{2 E_n({\bf q})},
\ee
with $a_n$ a dimensionless number.
In particular, all the $T=0$ contributions vanish as $\sim |{\bf q}|^4$
when ${\bf q}\to0$. By the Ward identities of the energy-momentum tensor, 
$ C_{33,33}(x_0,|{\bf q}|) = |{\bf q}|^{-4}\frac{d^4}{dx_0^4} C_{00,00}(x_0,{\bf q})$.
Noticing that for a scalar state $S$, $\<S| L^{-3/2}\int d^3{\bf x}\, \widehat T_{33}(x)|\Omega\>
=\frac{1}{3}\<S|L^{-3/2}\int d^3{\bf x}\, \widehat \theta(x)|\Omega\>$,
we then find the matrix element for the lightest glueball state,
\be
a_0 = \frac{s}{3M_0^3},
\ee
where $s$, the matrix element of the trace anomaly,
 is defined and calculated in~\cite{Meyer:2008tr}. Similarly,
realizing  that 
$C_{33,33}(x_0,{\bf 0},T) = \frac{1}{9}C_{kk,ll}(x_0,{\bf 0},T)+\frac{4}{3} C_{12,12}(x_0,{\bf 0},T)$,
we find that the matrix element $a_2$ corresponding to the tensor glueball is related to 
the matrix element $t$ defined in~\cite{Meyer:2008tr} by 
\be
a_2 = 2\sqrt{3}\,\frac{t}{3M_2^3}.
\ee

\bibliography{/afs/lns.mit.edu/user/meyerh/BIBLIO/viscobib}
\bibliographystyle{JHEP}


\end{document}